\documentclass[a4paper]{jpconf}
\usepackage[sort&compress,square,numbers]{natbib}
\usepackage{graphicx}
\usepackage{amssymb,amsmath}
\usepackage{units}
\usepackage{sidecap}
\usepackage{subfig}
\usepackage{dsfont}

\def\ring#1{{\mathaccent'27 #1}}

\begin{document}
\title{Vacuum Cherenkov radiation for Lorentz-violating fermions}

\author{Marco Schreck}

\address{Departamento de F\'{i}sica, Universidade Federal do Maranh\~{a}o \\
65080-805, S\~{a}o Lu\'{i}s, Maranh\~{a}o, Brazil}

\ead{marco.schreck@ufma.br}

\begin{abstract}
The current paper summarizes the content of a talk given on vacuum Cherenkov radiation emitted by Lorentz-violating fermions that are described in the context of the Standard-Model Extension. The decay rate will be obtained for various sets of controlling coefficients that will subsequently be constrained using cosmic-ray data.
\end{abstract}

\section{Introduction}

Ordinary Cherenkov radiation occurs whenever a particle carrying an electric charge moves through an optical medium with a speed that is larger than the phase velocity of light in that medium. Under this condition, the particle generates a large number of electric dipoles that start radiating their wave trains in phase such that they interfere constructively. The consequence is coherent radiation of high intensity that can be detected far away from the source. The latter is called Vavilov-Cherenkov radiation or most of the time simply Cherenkov radiation named after the Russian experimental physicists Vavilov and Cherenkov that discovered it in 1934. Only three years later, Frank and Tamm explained the effect theoretically in the context of classical electromagnetism. Cherenkov, Frank, and Tamm shared the Nobel Prize of 1958, which could be an explanation for why the name Vavilov is often omitted unjustifiedly. Cherenkov radiation is visible as a blue, glowing light in the water around nuclear reactors.

A Lorentz-violating background field turns the vacuum into an optical medium with nontrivial refractive index. The latter can depend on the energy of a propagating electromagnetic wave, which leads to dispersion. A direction dependence is possible, as well, and can result in birefringence, i.e., in a change of polarization of light as a function of propagation distance. Under certain circumstances, a charged particle loses energy by radiating photons when it moves through such a background field. Due to the fundamental characteristics shared with ordinary Cherenkov radiation, this process is referred to as vacuum Cherenkov radiation.

Vacuum Cherenkov radiation was probably mentioned in \cite{Beall:1970rw} for the first time in the context of modified static gravitational couplings. The process had been forgotten long since, until it had a revival at the end of the past millennium due to the rising interest in the field of Lorentz-symmetry violation. First, this increasing interest can be traced back to the emergence of a series of papers demonstrating how violations of Lorentz invariance can occur in quantum-gravity scenarios \cite{Kostelecky:1988zi,Gambini:1998it}. Second, the construction of the minimal Standard-Model Extension (SME) as a model-independent framework parameterizing all deviations from Lorentz invariance in the Standard Model and General Relativity in a coordinate-invariant and gauge-invariant way \cite{Colladay:1998fq,Kostelecky:2003fs} finally allowed for a systematic description of this effect. Since then, plenty of papers studying vacuum Cherenkov radiation have appeared with a small excerpt of this list given by \cite{Lehnert:2004be,Kaufhold:2007qd,Altschul:2006zz,Klinkhamer:2008ky}. Most of these works were carried out in the framework of either a modified electrodynamics or for modified photons where only recently papers have been published on gravitational Cherenkov radiation \cite{Kostelecky:2015dpa} and vacuum Cherenkov radiation for pions \cite{Altschul:2016ces} and for a modified electroweak sector \cite{Colladay:2017qfr}. Until lately, vacuum Cherenkov radiation for Lorentz-violating fermions has evaded investigation. This was the motivation for the article \cite{Schreck:2017isa} that is on the focus.

\section{Fermion sector of the minimal Standard-Model Extension}

Within the minimal SME, Lorentz-violating fermions are described by including all power-counting renormalizable Lorentz-violating terms into the Dirac Lagrange density, cf.~Eq.~(1) in~\cite{Kostelecky:2013rta}. Each of these contributions is decomposed into controlling coefficients and a field operator appropriately contracted such that the term is an observer Lorentz scalar. The controlling coefficients can be interpreted as background fields permeating the vacuum and resulting in velocity- and direction-dependent physics. The $b$, $d$, $g$, and $H$ coefficients are spin-nondegenerate, i.e., they produce two distinct dispersion relations for particles and antiparticles, respectively. The dispersion relation depends on the spin projection with respect to the quantization axis.

\section{Modified kinematics}

\begin{figure}
\begin{minipage}[b][3.5cm][c]{0.65\textwidth}
\subfloat[]{\label{fig:mass-shell-spin-degenerate}\includegraphics[scale=0.3]{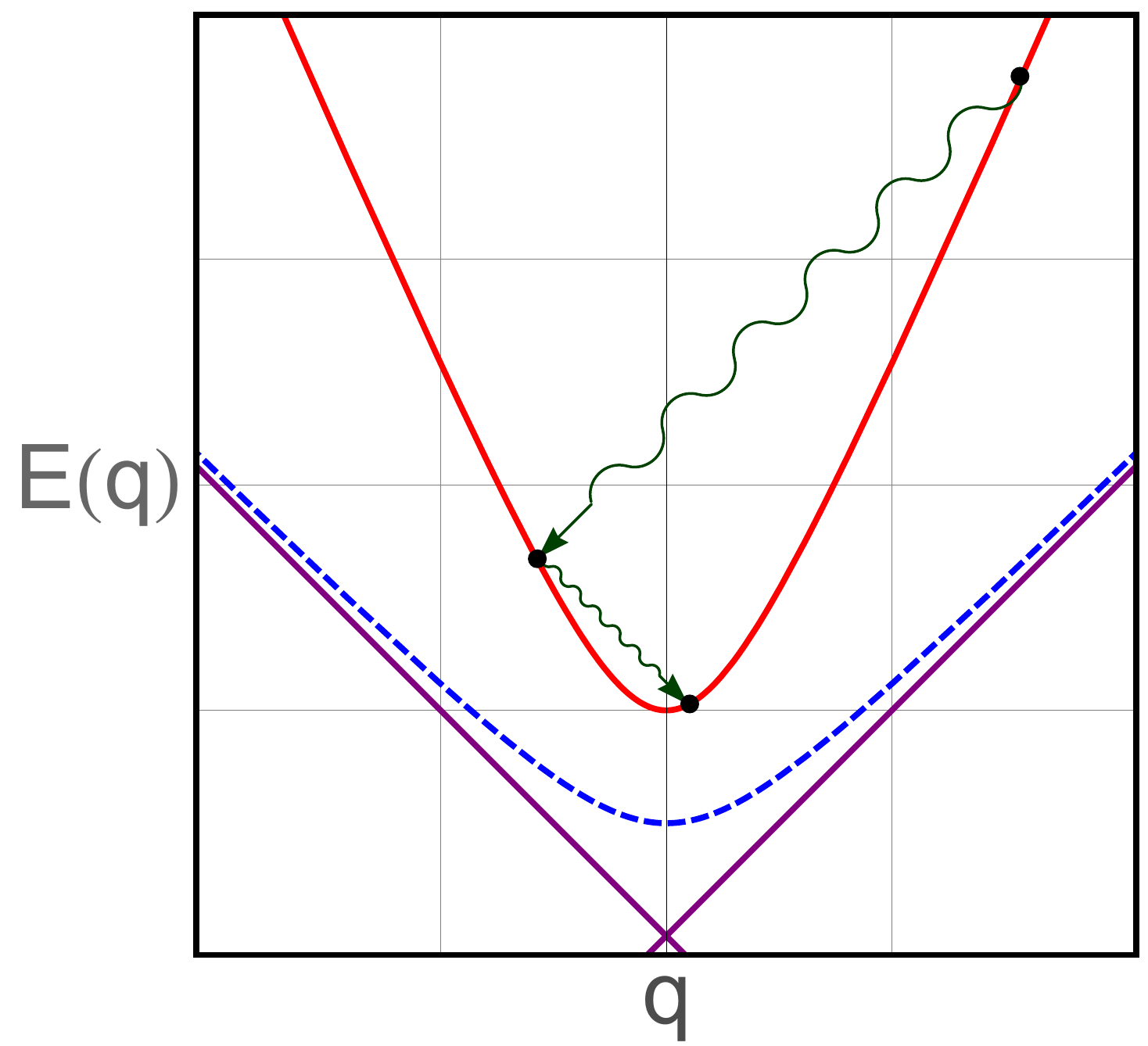}}\qquad
\subfloat[]{\label{fig:mass-shell-spin-nondegenerate}\includegraphics[scale=0.3]{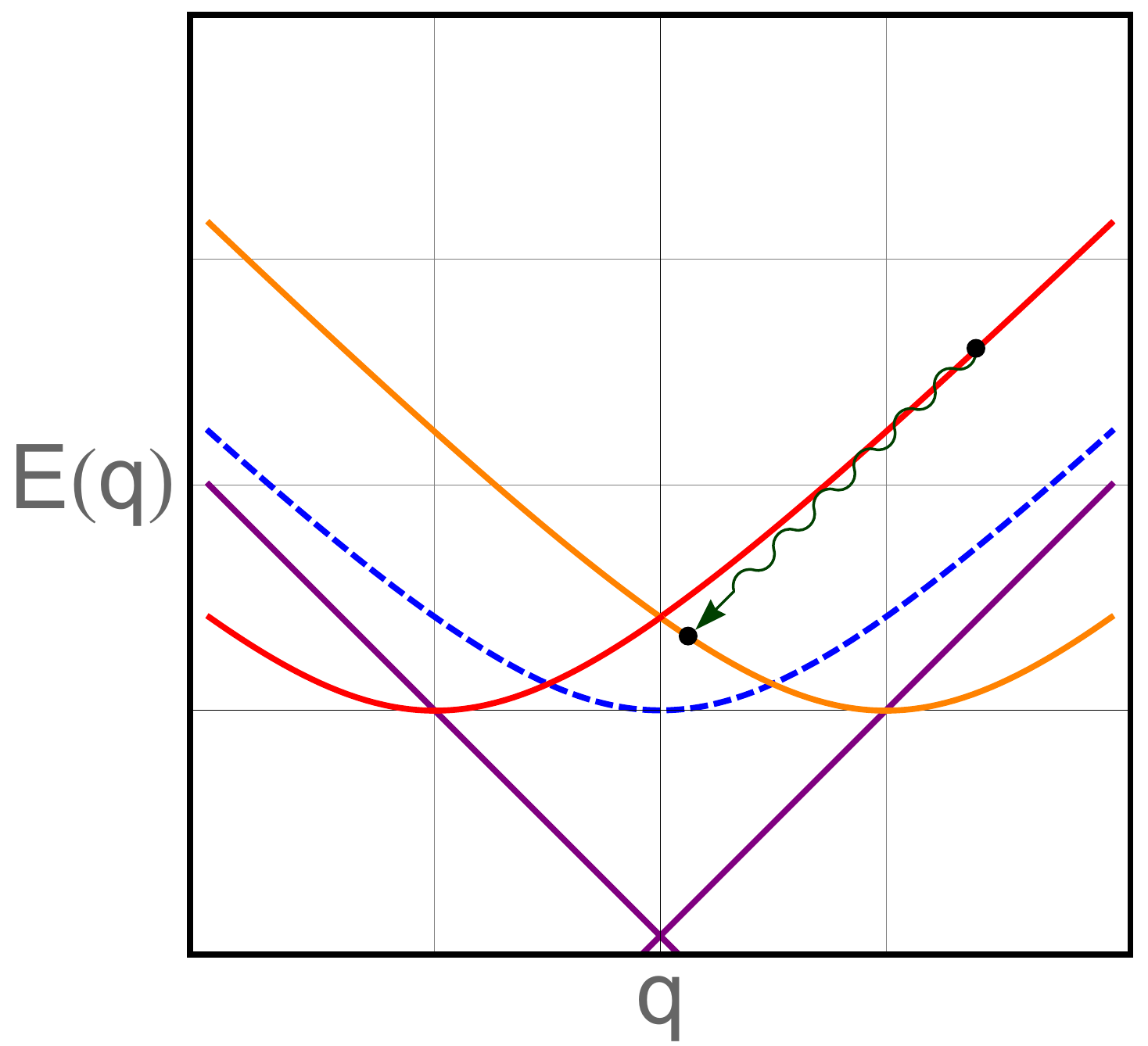}}
\end{minipage}
\begin{minipage}[b][3.5cm][b]{0.35\textwidth}
\caption{\label{label}\small Standard null cone (violet), standard mass shell (dashed, blue) and modified mass shell(s) (red, orange) for the spin-degenerate case with $\ring{c}=-1/2$ \protect\subref{fig:mass-shell-spin-degenerate} and for the spin-nondegenerate case with $\ring{b}=1$ \protect\subref{fig:mass-shell-spin-nondegenerate}. Photon emission is indicated by green, wiggly lines.}
\end{minipage}
\end{figure}

Basically, there are two different types of processes. The first involves an incoming fermion of a certain spin projection that emits a photon such that the spin projection is conserved. It is reasonable to study its kinematics based on plots of modified mass shells and null cones. As an example, we analyze the isotropic coefficient $\ring{c}$, cf.~Fig.~\ref{fig:mass-shell-spin-degenerate}. The modified mass shell evidently has a larger curvature in comparison to the standard mass shell. Considering a point on the mass shell far enough away from the origin, this property allows the modified fermion to emit a photon, lose energy, and reverse the direction of its momentum. An emitted photon corresponds to a line parallel to the standard null cone. The emission of several photons is possible as long as the fermion energy is large enough whereupon its energy decreases steadily and the direction of its momentum reverses for each emission. Using energy conservation, it is shown that the process is forbidden for the dimensionful coefficients, i.e., for $a$, $b$, and $H$.

The second type of process can only occur once the fermion is modified by spin-nondegenerate coefficients. Here, the spin projection of the initial fermion changes after photon emission, i.e., a spin flip occurs. In the literature, such a process is often called ``helicity decay.'' As an example, we consider an isotropic framework with a nonzero $\ring{b}$, cf.~Fig.~\ref{fig:mass-shell-spin-nondegenerate}. There are two different branches of the mass shell that intersect themselves and this intersection is visible as a cusp. With the fermion being on a certain branch, a photon can be emitted such that the final state lies on the other branch. This change of the branch corresponds to a spin flip. Helicity decays have certain characteristics that are absent for the previous type of process. First, multiple photons cannot be emitted, as this would require a further impossible change of the branch. Second, the direction of the fermion momentum is not reversed. After all, the required change of helicity for photon emission is induced by the spin flip. Energy conservation shows that helicity decays are allowed for all fermion coefficients except $a$. Hence, the latter coefficients are the only ones that do not render vacuum Cherenkov radiation possible.

\section{Modified dynamics}

The great success of the SME rests on the fact that it is based on field theory, i.e., it does not only allow for describing a modified kinematics, but enables studies of a modified dynamics, as well. This possibility is granted by the modified Lagrange density and Dirac equation. The latter can be solved for the dispersion relations and the spinor solutions describing asymptotic fermions in field theory. These solutions were derived in \cite{Reis:2016hzu} for various sets of both minimal and nonminimal coefficients, which is why that paper can be considered as the theoretical base for the work \cite{Schreck:2017isa}. The spinor solutions are needed to calculate the matrix element squared at tree level with an initial fermion of spin $s$ and momentum $q$ to a final fermion of spin $s'$ and momentum $p=q-k$:
\begin{subequations}
\begin{align}
|\mathcal{M}^{(s,s')}|^2&=4\pi\alpha\,\mathrm{Tr}[\Lambda^{(s)}(q)\Gamma^{\mu}\Lambda^{(s')}(q-k)\Gamma^{\nu}]\Pi_{\mu\nu}(k)\,, \\[2ex]
\Pi_{\mu\nu}(k)&\equiv \sum\nolimits_{\lambda=1,2} \varepsilon_{\mu}^{(\lambda)}(k)\varepsilon_{\nu}^{(\lambda)}(k)\,, \\[2ex]
\Gamma^{\nu}&=\gamma^{\nu}+c^{(4)\mu\nu}\gamma_{\mu}+d^{(4)\mu\nu}\gamma^5\gamma_{\mu}+e^{(4)\nu}\mathds{1}_4+\mathrm{i}f^{(4)\nu}\gamma^5+(1/2)g^{\varrho\sigma\nu}\sigma_{\varrho\sigma}\,, \\[2ex]
\Lambda_{ab}^{(s)}(q)&\equiv u_a^{(s)}(q)\bar{u}_b^{(s)}(q)\,,
\end{align}
\end{subequations}
with the fine-structure constant $\alpha=e^2/(4\pi)$, the unit matrix $\mathds{1}_4$ in spinor space, the chiral Dirac matrix $\gamma^5=\mathrm{i}\gamma^0\gamma^1\gamma^2\gamma^3$, and $\sigma^{\mu\nu}=(\mathrm{i}/2)[\gamma^{\mu},\gamma^{\nu}]$ where $\gamma^{\mu}$ are the standard Dirac matrices. Furthermore, $\Pi_{\mu\nu}$ is the sum over the physical photon polarization tensors with the polarization vectors $\varepsilon_{\mu}^{(\lambda)}$ for a transverse polarization mode $\lambda$. The trace in spinor space is carried out over a train of quantities. The minimal coupling of the modified Dirac theory to photons produces a modified vertex $\Gamma^{\nu}$ that involves all controlling coefficients contracted with a derivative in the Lagrange density. The quantities $\Lambda^{(s)}$ are matrices formed from a particle spinor $u^{(s)}$ and its Dirac-conjugated counterpart $\overline{u}^{(s)}\equiv u^{(s)\dagger}\gamma^0$. It was also demonstrated in \cite{Reis:2016hzu} that these matrices can be neatly computed directly from the modified propagators of the theory. The matrix element squared shown above is the field theoretical object needed to derive the decay rate for a helicity decay. However, once spin-degenerate coefficients are involved, the significant object is the following:
\begin{equation}
|\mathcal{M}|^2=\frac{1}{2}\sum\nolimits_{s,s'=\pm} |\mathcal{M}^{(s,s')}|^2\,.
\end{equation}
As we consider the initial particle beam to be unpolarized, the initial spin $s$ is averaged over. Since the final-particle state is unclear, the two spin projections $s'$ must be summed over. Finally, we also take into account Lorentz-violating contributions that contain additional time derivatives. These derivatives are known to spoil the proper time evolution of asymptotic fermion states, which is why they must be removed by a suitable transformation in spinor space~\cite{Colladay:2001wk}.

The matrix element squared delivers the probability for a certain process to occur with explicitly given final-particle momenta. However, quantum physics makes the final-particle state undetermined, which is why all possible momentum configurations have to be integrated over. This integration is six-dimensional and can immediately be reduced to three dimensions on the base of momentum conservation. The remaining three integrations must be performed by taking into account energy conservation. For an isotropic framework, we can parameterize the photon phase space with spherical polar coordinates and the azimuthal angle can be integrated over directly. The integration over the polar angle sets this angle to a particular value that is in accordance with energy conservation. The final remaining integration must be carried out over the magnitude of the photon momentum that runs from a vanishing momentum to a finite maximum $k_{\mathrm{max}}$. In the Lorentz-invariant case, the latter $k_{\mathrm{max}}$ vanishes, as there is no phase space available for this process to occur.

\section{Decay rates and threshold momenta}

\begin{figure}
\centering
\subfloat[]{\label{fig:decay-rates-vacuum-cherenkov-spin-conserving}\includegraphics[scale=0.36]{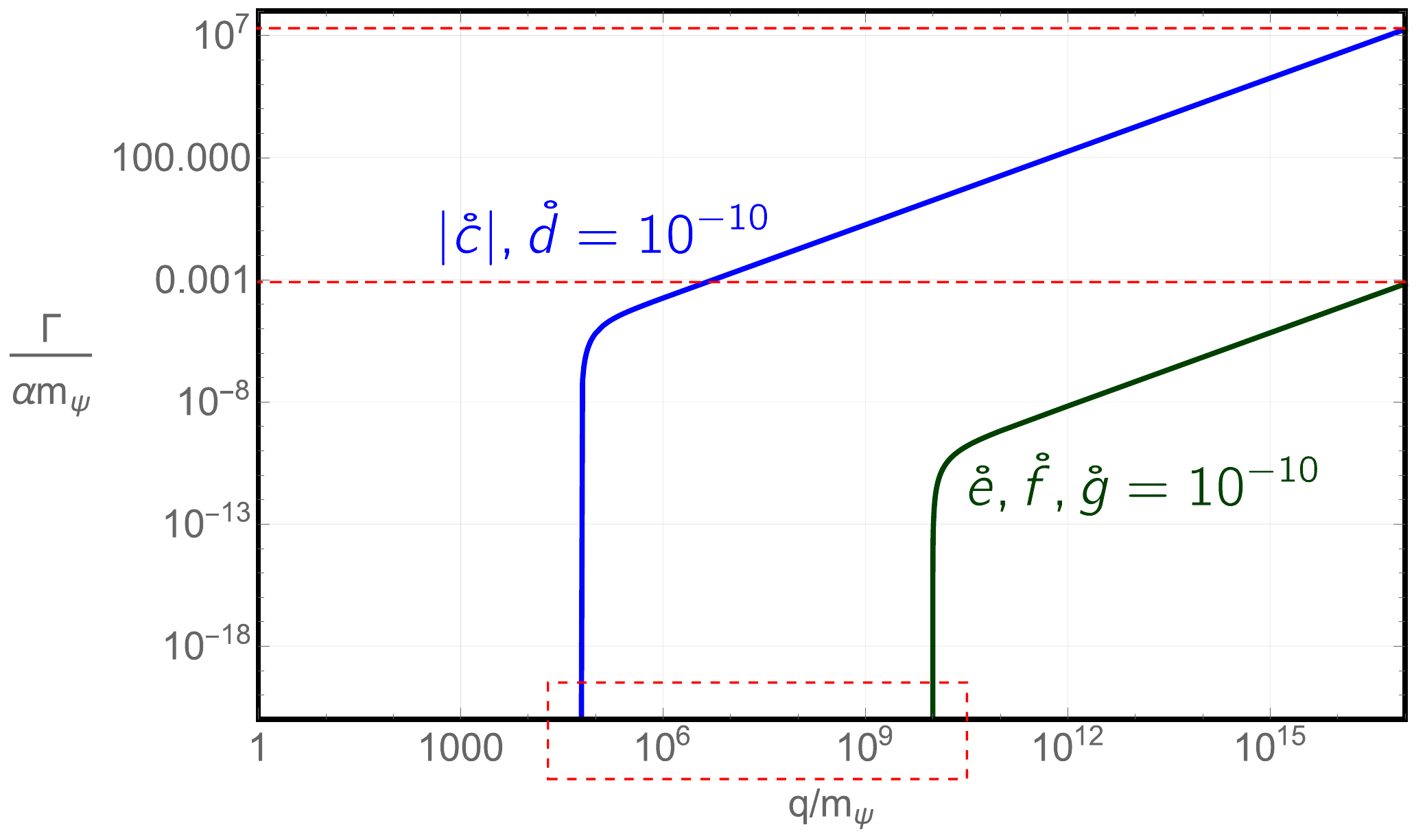}}\quad
\subfloat[]{\label{fig:helicity-decay-b-isotropic}\includegraphics[scale=0.36]{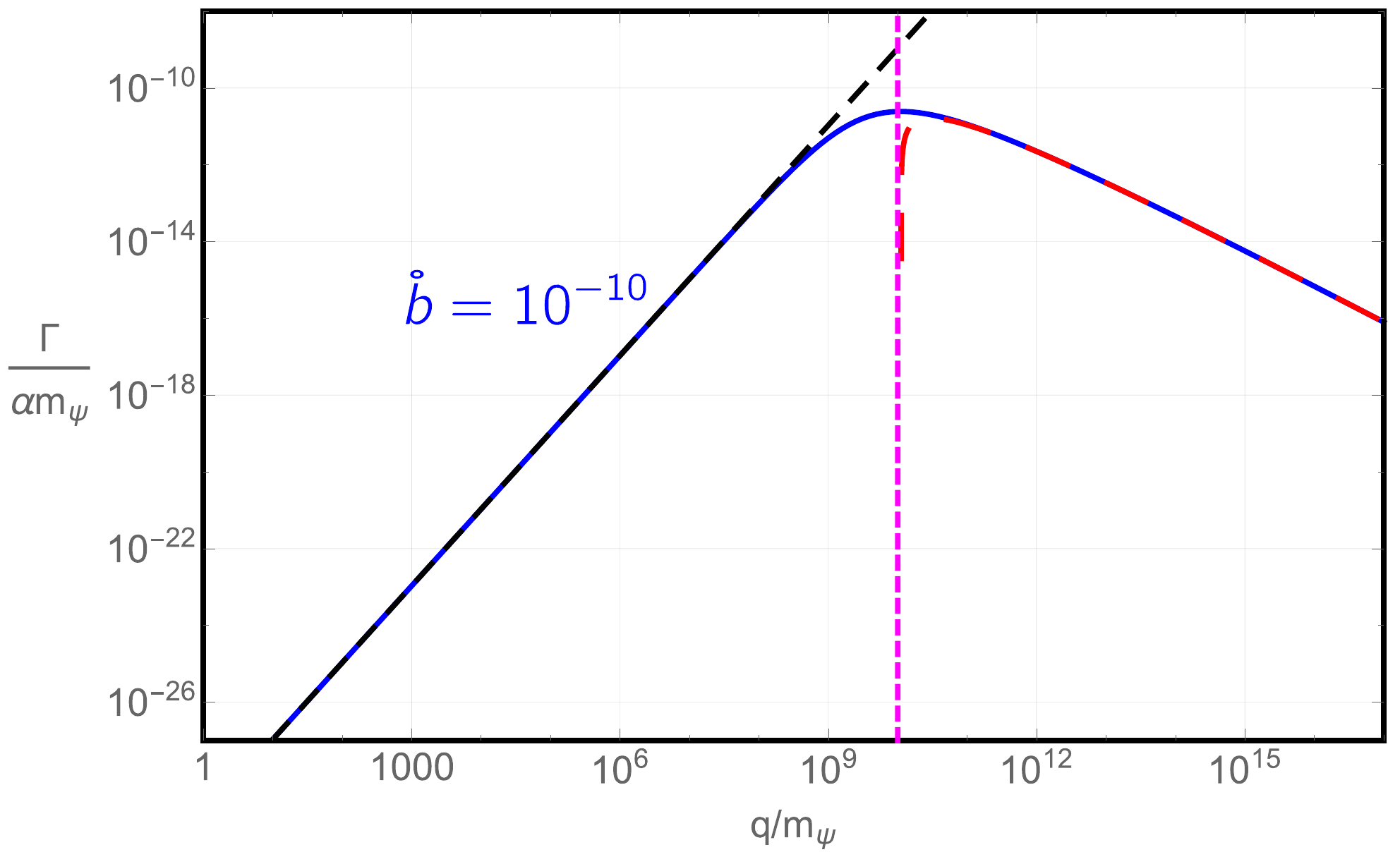}}
\caption{\label{fig:decay-rate}\small Double-logarithmic plot of decay rates $\Gamma/(\alpha m_{\psi})$ for vacuum Cherenkov radiation as functions of $q/m_{\psi}$ without a spin flip \protect\subref{fig:decay-rates-vacuum-cherenkov-spin-conserving} and with a spin flip of the initial fermion \protect\subref{fig:helicity-decay-b-isotropic}.}
\end{figure}

The explicit result for the matrix element squared is quite complicated even for simple coefficient choices. Hence, the phase space is integrated over numerically with computer algebra. For the spin-degenerate, isotropic coefficients, the results are shown in Fig.~\ref{fig:decay-rates-vacuum-cherenkov-spin-conserving}. The dimensionless coefficients are chosen to be $10^{-10}$ for the plot. For such tiny coefficients, the curves for $\ring{c}$, $\ring{d}$ and $\ring{e}$, $\ring{f}$, $\ring{g}$ cannot be distinguished from each other. An important property of the spin-conserving process is that it has a threshold, i.e., the decay rate goes to zero when the particle momentum approaches a certain value from above. Furthermore, for large momenta, the decay rate increases linearly with momentum. The decay rates for the second set of coefficients are suppressed in comparison to the decay rates for $\ring{c}$, $\ring{d}$ by an additional power in Lorentz violation.

It is also evident that the threshold energies for both groups of coefficients differ from each other. At leading order in Lorentz violation, the threshold momenta are given by
\begin{equation}
q^{\mathrm{th}}_{\ring{c}}=\frac{1}{2}\sqrt{\frac{3}{2}}\frac{m_{\psi}}{\sqrt{-\ring{c}}}\,,\quad q^{\mathrm{th}}_{\ring{d}}=\frac{1}{2}\sqrt{\frac{3}{2}}\frac{m_{\psi}}{\sqrt{\pm\ring{d}}}\,,\quad
q^{\mathrm{th}}_{\ring{e}}=\frac{m_{\psi}}{\pm\ring{e}}\,,\quad q_{\ring{g}}^{\mathrm{th}}=\frac{m_{\psi}}{\pm\ring{g}}\,.
\end{equation}
The threshold momenta are proportional to the fermion mass due to dimensional reasons. They depend on the inverse of the coefficients, which indicates that the process is forbidden for vanishing Lorentz violation. Furthermore, the coefficient $\ring{c}$ must be chosen negative to render the decay possible. For the spin-nondegenerate $\ring{d}$, $\ring{g}$ both signs of coefficients are allowed, as there are two dispersion relations whose first-order terms in Lorentz violation simply differ by a sign. Last but not least, both signs are also permissible for $\ring{e}$ although the latter is spin-degenerate. The lower signs for $\ring{d}$, $\ring{e}$, and $\ring{g}$ are always understood to hold for negative SME~coefficients.

The decay rates for the helicity decays of the dimensionful isotropic coefficients have completely different properties and for comparison, the result for a nonzero $\ring{b}$ is presented in Fig.~\ref{fig:helicity-decay-b-isotropic}. Here, the decay does not have a threshold, but it is highly suppressed by Lorentz violation. Furthermore, the rate has a maximum beyond which it decreases again, as for large momenta it becomes more and more difficult for the fermion spin to flip.

\section{Phenomenology}

We consider an ultra-high energetic particle propagating to the Earth after having been accelerated, e.g., in the shock front of a supernova remnant. At least for the $c$, $d$ coefficients, the radiated-energy rates are large enough such that the particle loses its energy fraction above the threshold after few centimeters, i.e., the decay is very efficient for these coefficients. Hence, when a cosmic ray is detected on Earth, its energy can be assumed to be smaller than the threshold.

To obtain constraints, we need a cosmic-ray event with known energy observed on Earth. A suitable event of $E=\unit[212]{EeV}$ was detected by the Pierre-Auger observatory. A conservative estimate is obtained by assuming that the primary was an iron nucleus with nucleon number $N=56$, i.e., each nucleon had an energy of $E/N$. We consider the quark structure of the nucleons where vacuum Cherenkov radiation can be emitted from the real up and down quarks only. A sophisticated analysis would require to take into account the parton distribution functions. However, we follow the simple, but reasonable estimate that each of these quarks carries a fraction $r=0.1$ of the total energy of the nucleon. We then obtain the following set of bounds on the up-quark coefficients from the threshold energies:
\begin{subequations}
\begin{align}
-3\times 10^{-23}&<\ring{c}^{\mathrm{u}}-(3/8)(\ring{f}^{\mathrm{u}})^2\,, \\[2ex]
-3\times 10^{-23}&<\ring{d}^{\mathrm{u}}<3\times 10^{-23}\,, \\[2ex]
-9\times 10^{-12}&<\ring{e}^{\mathrm{u}},\ring{g}^{\mathrm{u}}<9\times 10^{-12}\,.
\end{align}
\end{subequations}
Note that the $f$ coefficients squared can be moved to the $c$ coefficients by a field redefinition. The constraint on the $c$ coefficients is one-sided only, as vacuum Cherenkov radiation can only occur for a negative coefficient.

\section{Acknowledgments}

The author thanks the organizers of the workshop for the invitation and their hospitality. Furthermore, he acknowledges financial support via the grant FAPEMA/POS-GRAD/03978/15.

\newcommand{\newblock}{}

\end{document}